\documentclass[showpacs,preprintnumbers, amsmath,amssymb,superscriptaddress,preprint,floatfix]{revtex4}
\usepackage{graphicx}
\draft
\begin{document}
\newcommand {\bb}{\bibitem}
\newcommand {\be}{\begin{equation}}
\newcommand {\ee}{\end{equation}}
\newcommand {\bea}{\begin{eqnarray}}
\newcommand {\eea}{\end{eqnarray}}
\newcommand {\nn}{\nonumber}

\title{Scaling Relations in the Triplet Superconductor PrOs$_{4}$Sb$_{12}$}

\author{H. Won}

\address{Department of Physics, Hallym University,
Chuncheon 200-702, South Korea}

\author{S. Haas}
\author{D. Parker}
\author{K. Maki}

\address{Department of Physics and Astronomy, University of Southern
California, Los Angeles, CA 90089-0484 USA}

\pacs{74.25.Bt}

\date{\today}
\begin{abstract}

Scaling relations are one of the hallmarks of nodal superconductivity
since they contain information characteristic for gapless order parameters.
In this paper we derive the scaling relations for the 
thermodynamics and the thermal conductivity in the vortex state of the A and
B phases of the skutterudite PrOs$_{4}$Sb$_{12}$.  Experimental verification
of these scaling relations can provide further support for 
anisotropic gap functions which were previously considered for this 
material.

\end{abstract}
\maketitle
\noindent{\bf 1. Introduction}

Superconductivity in the filled skutterudite PrOs$_{4}$Sb$_{12}$ was
 discovered
in 2002 by Bauer et al \cite{1,2,3}, and has since
generated ever-increasing attention. In particular, 
the 
presence of at least two distinct phases, the A and B phase, in an applied
magnetic field is of great interest.  Experimentally, it was observed
that both phases have point nodes, and that the pairing channel appears
to be a triplet with chiral symmetry breaking. \cite{4,5,6}.  
However, the precise position of the
A-B phase boundary is still controversial.  For example, Measson et al 
\cite{7} found the A-B phase boundary to be almost parallel to H$_{c2}(T)$
of the A phase.  A possible explanation of this phase diagram
was recently proposed in terms of the gap functions \cite{6,8}
\bea
{\bf \Delta}_{A}({\bf k}) &=& {\bf d} e^{\pm i\phi_{i}}\frac{3}{2}(1-\hat{k}_{x}^{4}-\hat{k}_{y}^{4}-\hat{k}_{z}^{4}),\\
{\bf \Delta}_{B}({\bf k}) &=& {\bf d} e^{\pm i\phi_{3}}(1-\hat{k}_{z}^{4}).
\eea
Here  $e^{\pm i\phi_{1}}=( \hat{k}_{2} \pm i \hat{k}_{3})/\sqrt{\hat{k}_{2}^{2}+\hat{k}_{3}^{2}},
e^{\pm i \phi_{2}}=(\hat{k}_{3} \pm i\hat{k}_{1})/\sqrt{\hat{k}_{3}^{2}+\hat{k}_{1}^{2}}$,
and $e^{\pm i \phi_{3}}= (\hat{k}_{1}\pm i\hat{k}_{2})/\sqrt{\hat{k}_{1}^{2}+\hat{k}_{2}^{2}}$. 
The factor of 3/2 in the definition of
${\bf \Delta}_{A}({\bf k})$ ensures proper normalization of the angular dependence of the order parameter.
Furthermore, in Eq.(2) we choose the nodal direction to be parallel to [001],
because this p+h-wave order parameter symmetry is consistent with the
magnetothermal conductivity data of Izawa et al \cite{6}.

In 1997, Simon and Lee \cite{9} introduced scaling relations for d-wave
superconductors.  More recently, following Volovik's approach \cite{10} K\"{u}bert
and Hirschfeld \cite{11} obtained a scaling function for the 
quasiparticle density of states (DOS) in the vortex state of d-wave
superconductors.  This expression for the DOS contains the scaling relations of the thermodynamic response functions as well as
the thermal conductivity \cite{11,12,13,14,15}.  From their very general derivation
it is clear that such scaling laws must apply to all nodal superconductors 
which have a comparable
low-energy quasiparticle DOS G(E)$\sim |E|/\Delta$ for $|E| < 0.3 \Delta$
in the absence of a magnetic field.  If the above proposals (Eqs. (1)
and (2)) for 
$\Delta({\bf k})$ are correct, both phases of PrOs$_{4}$Sb$_{12}$ would 
fall into this category. 
Experimentally, scaling laws for the specific heat have been
verified experimentally in the cuprate superconductor YBCO \cite{16}, in the 
the ruthenate superconductor Sr$_{2}$RuO$_{4}$\cite{17} with a 
magnetic field {\bf H} $\parallel$ [001], and in the thermal conductivity
of the heavy-fermion superconductor UPt$_{3}$ \cite{18}.

These measurements are consistent with the theory of scaling in nodal
superconductors.  Hence, scaling relations can be regarded as one of the
hallmarks of nodal superconductivity.  So far, however, scaling laws have 
only been studied in superconductors with line nodes, such as d-wave
and f-wave order parameters.
 The object of this work is to extend these early
analyses to superconductivity with point nodes by focusing
on the skutterudite compound 
PrOs$_{4}$Sb$_{12}$. \cite{27}.  

\noindent{\bf 2. Quasiparticle Density of States}

Let us first consider the quasiparticle density of states 
in this compound, using the gap functions for the
A and B phases given by Eqs. (1) and (2).  
In the absence of a magnetic field, the low-energy quasiparticle DOS can
then be approximated by \cite{6,8}
\bea
G_{A}(E) &=& \frac{\pi}{4}|E|/\Delta, \\
G_{B}(E) &=& \frac{\pi}{8}|E|/\Delta.
\eea
These equations are 
accurate in the low-energy
regime $E <0.3 \Delta$.  Furthermore,
in the vortex state the effect of the supercurrent can be introduced
by letting $E \rightarrow E - {\bf v}\cdot{\bf q}$, where ${\bf v}\cdot{\bf q}$ denotes the Doppler shift.  Following the derivation of K\"{u}bert and
Hirschfeld \cite{11} we obtain
\bea
G_{A}(E,{\bf H}) &=& \frac{v\sqrt{eH}}{6\Delta}\sum_{i=1}^{3}\sin\theta_{i}
g(E/\epsilon_{i}), \\
G_{B}(E,{\bf H}) &=& \frac{1}{2\Delta}\epsilon_{3}g(E/\epsilon_{3}).
\eea
Here, the scaling function is given by
\bea
g(s)&=& \frac{\pi}{4}s(1+\frac{1}{2s^{2}}), s > 1 \\
    &=& \frac{3}{4}\sqrt{1-s^{2}}+\frac{1}{4s}(1+2s^{2})\arcsin(s),
\ \ \ \  s \leq 1
\eea
and
\bea
\epsilon_{i}&=& \frac{v}{2}\sqrt{eH}\sin\theta_{i}, \\
\sin\theta_{1}&=& (1-\sin^{2}\theta\cos^{2}\phi)^{1/2}\\
\sin\theta_{2}&=& (1-\sin^{2}\theta\sin^{2}\phi)^{1/2}\mathrm{\,\,and\,}\\
\sin\theta_{3} &=&  \sin\theta
\eea
$(\theta,\phi)$ are the angles indicating the direction 
of the applied magnetic field {\bf H}.  Note that in these
equations the effect of impurity scattering is neglected. Therefore
this result is valid only in the superclean
limit, i.e., $(\Gamma\Delta)^{1/2} < |E|, \epsilon < \Delta$, where 
$\Gamma$ is the quasiparticle scattering rate of the normal state.\cite{22} 
In order to observe scaling behavior it thus appears necessary to have
$\Gamma \leq 0.01\Delta$.  If such a sample is available
the DOS obtained above should then be
accessible by scanning tunneling microscope measurements.  As seen from Eqs.(5) and (6), 
both G$_{A}(E,{\bf H})$ and G$_{B}(E,{\bf H})$ obey scaling laws.  In
particular the scaling law for G$_{B}(E,{\bf H})$ is the same as in d-wave
superconductors.  

\begin{figure}[h]
\includegraphics[width=15.5cm]{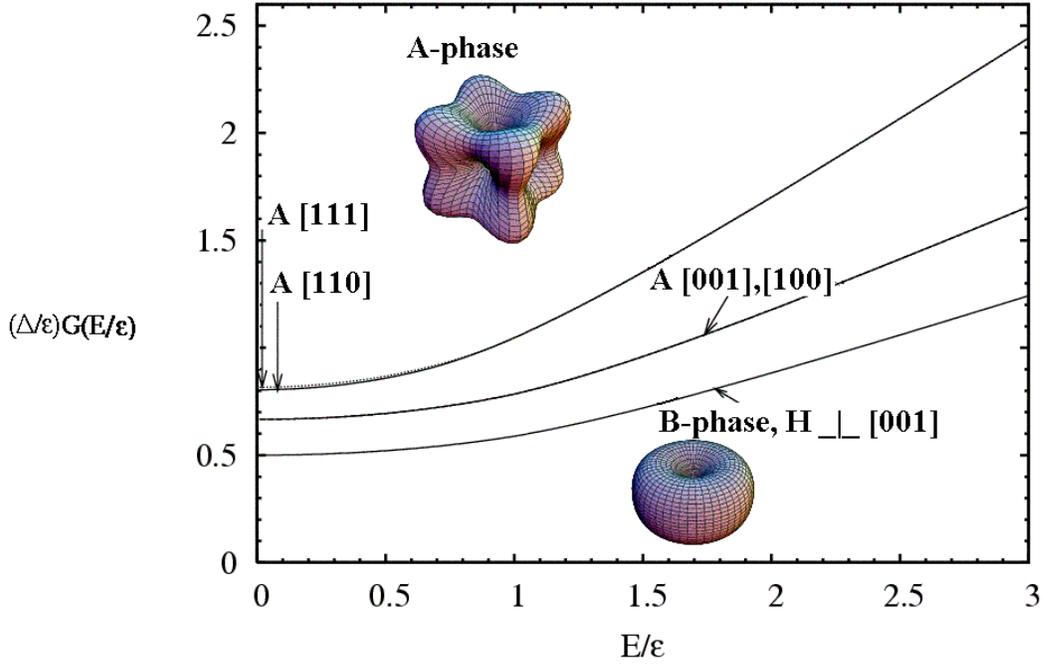}
\caption{The functions G$_{A}(E,{\bf H})$ and G$_{B}(E,{\bf H})$ 
along various directions of the applied magnetic field.}
\end{figure}

In Fig. 1 G$_{A}(E,{\bf H})$ is shown for ${\bf H} \parallel [1 1 1], 
{\bf H} \parallel [110]$ and ${\bf H} \parallel [1 0 0]$ and G$_{B}(E,{\bf H})$
for ${\bf H} \perp [0 0 1]$.  Along specific field directions we obtain
\bea
G_{A}(E,{\bf H}) &=& \frac{2\epsilon}{3\Delta}g\left(\sqrt{\frac{3}{2}}E/\epsilon\right), 
\mathrm{\,for \,\,{\bf H}  \parallel [1 1 1]}, \\
&=& \frac{\epsilon}{3\Delta}(g(\sqrt{2}E/\epsilon)+g(E/\epsilon)), 
\mathrm{\,for \,\, {\bf H}  \parallel [1 1 0]}, \\
&=& \frac{2\epsilon}{3\Delta}g(E/\epsilon), 
\mathrm{\,for \,\, {\bf H}  \parallel [1 1 0]},
\eea
where $\epsilon=\frac{v}{2}\sqrt{eH}$.  Note that G$_{A}(E,{\bf H})$ for 
${\bf H}  \parallel [111]$ and ${\bf H}  \parallel [1 1 0]$ look very similar.
Also due to the cubic symmetry of $|\Delta({\bf k})|$ in the A phase
(see the insert in Fig. 1) the cases ${\bf H} \parallel [1 1 1]$,
${\bf H}  \parallel [-1 1 1]$ and ${\bf H}  \parallel [1 -1 1]$ etc.
are equivalent.

In the B phase, the specific heat, the spin susceptibility,
the superfluid density, and
the nuclear spin lattice relaxation rate are then given by 
\bea
C_{s}(T,{\bf H})/\gamma_{S}T &=& 
\frac{1}{2}\frac{\epsilon}{\Delta}f(T/\epsilon) \\
\chi_{s}(T,{\bf H}) &=& \frac{1}{2}\frac{\epsilon}{\Delta}h(T/\epsilon) \\ 
\rho_{s \parallel}(T,{\bf H})/\rho_{s}(0,0) &=& 1 - \frac{3\epsilon}{2\Delta}
h(T/\epsilon) \\ 
T_{1}^{-1}(T,H)/T_{1N}^{-1}&=& (\frac{\epsilon}{2\Delta})^{2} J(T/\epsilon),
\eea
where $\epsilon=\epsilon_{3}$ and $\rho_{s \parallel}$ denotes the current
parallel to the nodes (i.e. {\bf J} $\parallel [001]$). This expressions
contain further scaling functions, 
\bea
f(T/\epsilon) &=& \frac{3}{2\pi^{2}}(\frac{\epsilon}{T})^{3} \int_{0}^{\infty}
ds \, s^{2}g(s) \mathrm{sech}^{2}\left(\frac{\epsilon s}{2T}\right) \\
h(T/\epsilon) &=& \frac{\epsilon}{2T} \int_{0}^{\infty}
ds \,g(s) \mathrm{sech}^{2}\left(\frac{\epsilon s}{2T}\right)\mathrm{and\,\,}\\
J(T/\epsilon) &=& \frac{\epsilon}{2T} \int_{0}^{\infty}
ds \, s \,g^{2}(s) \mathrm{sech}^{2}\left(\frac{\epsilon s}{2T}\right)
\eea
These expressions can be expanded in the low-temperature and
high-temperature limits, with asymptotics given by
\bea
f(T/\epsilon)&=& 1 + \frac{7\pi^{2}}{30}(T/\epsilon)^{2} + \ldots, 
\mathrm{\, for \ \   T/\epsilon \ll 1} \\
&=& \frac{27\zeta(3)}{4\pi}\frac{T}{\epsilon}+\frac{3}{4\pi}\ln(2)
\frac{\epsilon}{T} + \ldots \mathrm{\,\,for\,\, \frac{T}{\epsilon} \gg 1}\\
h(T/\epsilon)&=& 1+ \frac{\pi^{2}}{18}(T/\epsilon)^{2}+ \ldots, 
\mathrm{for\,\, T/\epsilon \ll 1} \\
&=& \frac{\pi\ln2}{2}\frac{T}{\epsilon}+\frac{\pi}{32}\frac{\epsilon}{T}
\ln \left(1+(\frac{2T}{\epsilon})^{2}\right) + \ldots, \mathrm{\,\,for\, \frac{T}{\epsilon} \gg 1}\\
J(T/\epsilon)&=& 1 + \frac{\pi^{2}}{9}\left(\frac{T}{\epsilon}\right)^{2} +\ldots, 
\mathrm{\,for \, T/\epsilon \ll 1} \\
&=& (\frac{\pi}{4})^{2}\left(\frac{1}{3}(\frac{\pi T}{\epsilon})^{2}\right) 
+\ldots, 
\mathrm{for \frac{T}{\epsilon} \gg 1}
\eea
These scaling 
functions are the same as in d-wave superconductors \cite{14} and are 
shown in Fig. 2, where we introduced 
$F= f - \frac{27\zeta(3)T}{4\pi\epsilon}$, $K = h - 
\frac{\pi\ln 2}{2}\frac{T}{\epsilon}$ and $ G = J - \frac{\pi^{4}}{48}
(\frac{T}{\epsilon})^{2}$.

\begin{figure}[h]
\includegraphics[width=8cm]{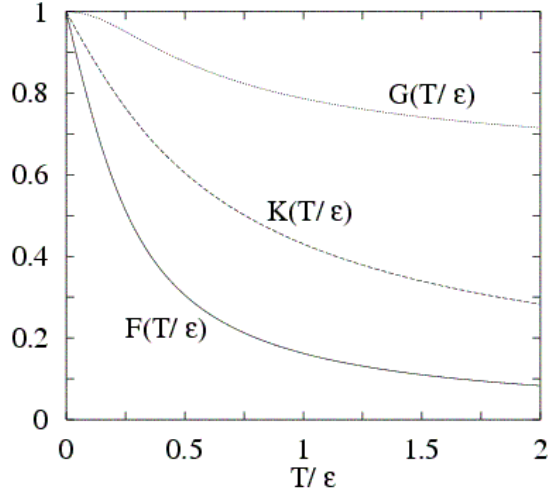}
\caption{The scaling
functions F(T/$\epsilon$), G(T/$\epsilon$), and K(T/$\epsilon$).}
\end{figure}

In analogy,
for the A phase Eqs. (16), (17) and (18) are replaced by \cite{28} 
\bea
C_{s}(T,{\bf H})/\gamma_{S}T &=& 
\frac{1}{3\Delta}\sum_{i=1}^{3}\epsilon_{i}f(T/\epsilon_{i}) \\
\chi_{s}(T,{\bf H}) &=& \frac{1}{3\Delta}\sum_{i=1}^{3}
\epsilon_{i}h(T/\epsilon_{i})\\ 
\rho_{s \parallel}(T,{\bf H})/\rho_{s}(0,0) &=& 1 - \frac{1}{3\Delta}
\sum_{i=1}^{3}\epsilon_{i}h(T/\epsilon_{i}) 
\eea

\noindent{\bf 3. Thermal Conductivity}

In order to determine the thermal conductivity it is necessary
to include the effect of impurity scattering because unlike the
thermodynamic response functions treated in the previous section, the scaling function for the thermal conductivity depends on the strength of the disorder, 
i.e. whether the
impurity scattering is in the Born limit or the unitary limit \cite{12}.
As we shall see below, the scaling functions for PrOs$_{4}$Sb$_{12}$
turn out to be particularly simple if the heat current is parallel
to a pair of point nodes.  On the other hand, in
the B phase the heat current has to be parallel to the nodal directions, in
order to see an appreciable heat current.  It appears that
this condition is realized experimentally as reported 
in \cite{6}.  
Otherwise the thermal conductivity would be much smaller since
it vanishes like 
$T(\frac{T}{\Delta})^{2}$ as T approaches zero.  In the A phase the heat
current is always appreciable, although the thermal conductivity loses the 
cubic symmetry in the vortex state, unless {\bf H} is directed along
some symmetric direction (for example, ${\bf H} 
\parallel [111], [-111],$ etc.).  

Following the derivation of Refs. \cite{23,24}, the
thermal conductivity is given by
\bea
\kappa_{zz} &=& \frac{3n}{4mT^{2}}\int_{0}^{\infty}d \omega \, \omega^{2}
\left\langle\frac{z^{2}h(\omega, {\bf H})}{\tilde{\Gamma}(\omega, {\bf H})}
\right\rangle \mathrm{sech}^{2}(\omega/2T)
\eea
where
\bea
h &=& \frac{1}{2}\left(1 + \frac{|\tilde{\omega}-{\bf v}\cdot{\bf q}|^{2}-
\Delta^{2}f^{2}}{|(\tilde{\omega}-{\bf v}\cdot{\bf q})^{2}-\Delta^{2}f^{2}|}
\right)
\eea
and
\bea
\tilde{\Gamma}&=& \mathrm{Im} \sqrt{(\tilde{\omega}-{\bf v}\cdot{\bf q})^{2}-
\Delta^{2}f^{2}}
\eea
Here $< ... >$ denotes the averages over the Fermi surface and vortex lattice
\cite{13}.  In the superclean limit $((\Gamma\Delta)^{1/2} < \omega, |{\bf v} 
\cdot {\bf q}| < \Delta)\, \tilde{\omega}$ is given by
\bea
\tilde{\omega} &=& \omega + i\Gamma \left\langle\frac{|\tilde{\omega} -
{\bf v}\cdot {\bf q}|}{\sqrt{(\tilde{\omega}-{\bf v} \cdot {\bf q})^{2} -
\Delta^{2}f^{2}}}\right\rangle \\
& \simeq & \omega+i\Gamma G(\omega,{\bf H})
\eea
in the Born limit. And in the unitary limit we find 
\bea
\tilde{\omega} = \omega+i\Gamma G^{-1}(\omega,{\bf H})
\eea
where G($\omega,{\bf H}$) for the A and B phase have been
defined in Eqs. (5) and (6).

Let us first consider the Born limit.  Substituting Eq.(35) into Eq.(32)
we obtain
\bea
\kappa_{zz} &=& \frac{3n}{8mT^{2}\Gamma}\int_{0}^{\infty}d \omega \, \omega^{2}
\frac{<|\omega-{\bf v}\cdot{\bf q}|>^{'}}{<|\omega-{\bf v}\cdot{\bf q}|>}\mathrm{sech}^{2}(\omega/2T)
\eea
where
$<|\omega-{\bf v}\cdot{\bf q}|>^{'}$ is the same as $<|\omega-{\bf v}\cdot{\bf q}|>$ but with 
contributions from nodes at (001) and (00 -1) only.  Then in the B phase $<|\omega-{\bf v}\cdot{\bf q}|>^{'} =  <|\omega-{\bf v}\cdot{\bf q}|>\,\,$ and thus 
$\kappa^{B}_{zz} = \frac{3}{2}\kappa_{n}$, where $\kappa_{n}=
\frac{\pi^2 T n}{6m\Gamma}$ is the thermal conductivity in the normal state.
The thermal conductivity in the Born limit is independent of {\bf H}.  On
the other hand in the
A phase$\,\, <|\omega-{\bf v}\cdot{\bf q}|>^{'} \neq <|\omega-{\bf v}\cdot{\bf q}|>$, leading to
\bea
\kappa_{zz}^{A}/\kappa_{n}&=& \frac{1}{2}, \mathrm{\ \ \ for\,\,\, T \gg \epsilon}, \\
&=& \frac{3}{2} \epsilon_{3}/(\epsilon_{1}+\epsilon_{2}+\epsilon_{3}),
\mathrm{\,\,for\,\, T \ll \epsilon}.
\eea
The $\epsilon_{i}$'s were defined in Eq.(9).

The scaling law is of greater interest in the unitary limit.  First
let us consider the B phase where $<|\omega-{\bf v}\cdot{\bf q}|>^{'} =  <|\omega-{\bf v}\cdot{\bf q}|>$.  Substituting Eq. (37) into Eq. (32) one obtains 
\bea
\kappa_{zz}^{B} &=& \frac{3\pi^{2}n}{512m\Gamma(T\Delta)^{2}}\int_{0}^{\infty}\
d\omega \omega^{2}<|\omega-{\bf v}\cdot{\bf q}|>^{2}\mathrm{sech}^{2}(\omega/2T)\\
 &=& \frac{3n\epsilon^{5}}{32m\Gamma(T\Delta)^{2}}\int_{0}^{\infty}\
ds\, s^{2} g^{2}(s) \mathrm{sech}^{2}(\epsilon s/2T)
\eea
and the scaling function is defined as
\bea
F(T/\epsilon) \equiv \frac{\kappa_{zz}^{B}(T,{\bf H})}{\kappa_{zz}^{B}(T,0)} =
\frac{120}{\pi^6}(\frac{T}{\epsilon})^{-6}\int_{0}^{\infty}\
ds\, s^{2} g^{2}(s) \mathrm{sech}^{2}(\epsilon s/2T)
\eea
where $\epsilon=\epsilon_{3}$.  

\begin{figure}[h]
\includegraphics[width=8cm,angle=270]{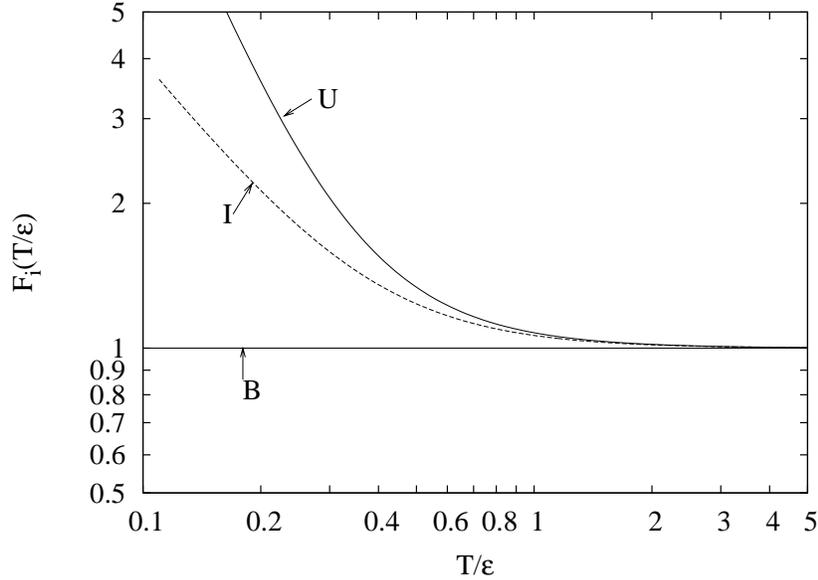}
\caption{The scaling function F(T/$\epsilon$) for the unitary limit (U),
the Born limit (B) and the case without inversion symmetry (I) are shown
as a function of T/$\epsilon$.}
\end{figure}

where $\epsilon=\epsilon_{3}$.  This scaling function is shown in Fig. 3,
with asymptotics given by
\bea
\kappa_{zz}^{B}/\kappa_{n} &=& \frac{21 \pi^4}{640}(\frac{T}{\Delta})^2\left(
1+ \frac{5}{7}(\frac{\epsilon}{\pi T})^2 + \ldots \right), \,\,
\mathrm{for\,\, \epsilon \ll T} \\
&=& \frac{3\pi^2}{8}(\frac{\epsilon}{\Delta})^2\left(1+ \frac{7\pi^2}{15}(\frac{T}{\epsilon})^2 + \ldots\right), \,\, \mathrm{for\, \epsilon \gg T}
\eea
and 
\bea
F^{B}(T/\epsilon) &=& 1+ \frac{5}{7}\left(\frac{\epsilon}{\pi T}\right)^2 + \ldots, \,\,
\mathrm{for \, \epsilon \ll T} \\
&=& \frac{80}{7\pi^{4}}(\frac{\epsilon}{\Delta})^2\left(1+ \frac{7\pi^2}{15}(\frac{T}{\epsilon})^2 + \ldots \right), \,\, \mathrm{for\, \epsilon \gg T}
\eea

This scaling function $F(T/\epsilon)$ is the same in other nodal 
superconductors such as those with d-wave symmetry.  For example,   
$F^{B}(T/\epsilon)$ describes very well the scaling behavior 
recently observed by Suderow et al \cite{18} in UPt$_{3}$.

In the A phase the scaling function is somewhat more complicated.  We find
\bea
\kappa_{zz}^{A}&=& \frac{n\epsilon_{3}}{72m\Gamma(\Delta T)^{2}}
\int_{0}^{\infty}\, d\omega \,\omega^{2} \, g(\omega/\epsilon_{3})
\left(\sum_{i=1}^{3}\epsilon_{i}g(\omega/\epsilon_{3})\right) \, \mathrm{sech}^{2}(\omega/2T)
\eea
In particular for ${\bf H} \parallel [1 1 1]$ and $[1 0 0]$ Eq.(48)
reduces to 
\bea
\kappa_{zz}^{A}&=& \frac{n\epsilon_{3}^5}{24m\Gamma(\Delta T)^{2}}
\int_{0}^{\infty}\, ds \,s^{2} \, g^{2}(s) \mathrm{sech}^{2}\left(\epsilon_{3}s/2T\right)
\mathrm{\,for\, {\bf H} \parallel [1 1 1], \,\,and} \\
&=& \frac{n\epsilon_{3}^5}{36m\Gamma(\Delta T)^{2}}
\int_{0}^{\infty}\, ds \,s^{2} \, g^{2}(s) \mathrm{sech}^{2}\left(\epsilon_{3}s/2T\right)
\mathrm{\,for \, {\bf H} \parallel [1 0 0]}
\eea

where $\epsilon_{3}=\frac{v\sqrt{eH}}{2}\sqrt{\frac{2}{3}}$ and
$\frac{v\sqrt{eH}}{2}$ for ${\bf H} \parallel [1 1 1]$ and 
${\bf H} \parallel [1 0 0]$ respectively.  Therefore in these two cases 
we will have the same scaling function as $F^{B}(T/\epsilon)$.

\noindent{\bf 4. Concluding Remarks}

We conclude that the scaling behavior of the universal heat conduction and the
thermal conductivity can be regarded as a hallmark of nodal 
superconductivity \cite{22}.  Moreover,
the scaling function $F^{B}(T/\epsilon)$ describes
the thermal conductivity data measured
in UPt$_{3}$ by Suderow et al \cite{18}
very well.  In this paper, we have found that the thermal conductivity in
both the A and B phases of PrOs$_{4}$Sb$_{12}$ exhibits a number of
characteristic
scaling relations.  The directional dependence of these scaling relations on
${\bf H}$ and ${\bf q}$ is expected to further confirm the nodal structure of
$\Delta({\bf k})$ proposed in \cite{6}.

In the course of the present study we have also observed that the scaling
behavior of the thermal conductivity in CePt$_{3}$Si found by Izawa et al
\cite{25,26} is very unusual.  Their data appears to be more consistent
with the case where the inversion symmetry of the impurity scattering is
broken.  Clearly, further study of scaling laws in
nodal superconductors will open a new point of view on the whole subject.

{\bf Acknowledgements}

We thank K. Izawa and Y. Matsuda for sharing with us unpublished data of
the thermal conductivity in CePt$_{3}$Si, which gave a strong motivation for
the present analysis.  HW
acknowledges support from the Korean Science and Engineering Foundation 
(KOSEF) through grant number R05-2004-000-10814, while SH acknowledges support
from the National Science Foundation through grant number DMR-0089882.

\end{document}